\documentclass[aps,prl,reprint,superscriptaddress]{revtex4-2}

\usepackage{graphicx}
\usepackage{siunitx}
\usepackage{caption}
\usepackage{lineno}
\usepackage{framed}
\usepackage{ragged2e}
\usepackage{multirow}
\usepackage{subcaption} % For subfigures
\begin{document}

\title{Integrated photonics for continuous-variable quantum optics}% 

\author{Rachel N. Clark}
\email[]{rachel.clark@bristol.ac.uk}
\affiliation{Quantum Engineering Technology Labs, H. H. Wills Physics Laboratory and Department of Electrical and Electronic Engineering, University of Bristol, Bristol, UK}

\author{Bethany Puzio}
\affiliation{Quantum Engineering Technology Labs, H. H. Wills Physics Laboratory and Department of Electrical and Electronic Engineering, University of Bristol, Bristol, UK}
\affiliation{Quantum Engineering Centre for Doctoral Training, H. H. Wills Physics Laboratory and Department of Electrical and Electronic Engineering, University of Bristol, Bristol, UK}

\author{Oliver M. Green}
\affiliation{Quantum Engineering Technology Labs, H. H. Wills Physics Laboratory and Department of Electrical and Electronic Engineering, University of Bristol, Bristol, UK}
\affiliation{Quantum Engineering Centre for Doctoral Training, H. H. Wills Physics Laboratory and Department of Electrical and Electronic Engineering, University of Bristol, Bristol, UK}

\author{Siva T. Pradyumna}
\affiliation{Department of Physics, University of Colorado, Boulder, Colorado 80305, USA}

\author{Oliver Trojak}
\affiliation{School of Physics and Astronomy, University of Southampton, Southampton SO17 1BJ, UK}

\author{Alberto Politi}
\affiliation{School of Physics and Astronomy, University of Southampton, Southampton SO17 1BJ, UK}

\author{Jonathan C. F. Matthews}
\affiliation{Quantum Engineering Technology Labs, H. H. Wills Physics Laboratory and Department of Electrical and Electronic Engineering, University of Bristol, Bristol, UK}

\date{\today}% It is always \today, today,

\begin{abstract}

Quantum technologies promise profound advances in communication security, sensing and computing. The underpinning hardware must be engineered to generate, manipulate and detect quantum phenomena with exceptional performance, whilst being mass-manufacturable for real-world applications. A leading approach is chip-scale quantum photonics. The continuous-variable regime for quantum optics has been exploited in a number of technologies, including the detection of gravitational waves, by operating below the standard quantum limit of the light's shot noise. The availability of room-temperature, deterministic sources and high efficiency detectors suitable for continuous-variable state generation and measurement is a compelling motivation for this particular paradigm. %Integrating these experiments on-chip provides a potential blueprint for scalable photonic technologies. 
This review focusses on efforts to integrate sources and detectors of continuous-variable light states into chip-scale photonic integrated circuits.
\end{abstract}


\maketitle

\section{Introduction}
\label{sec:Introduction}
%\linenumbers
Photonics is a leading approach to develop quantum technologies, with chip-scale photonic integrated circuits (PICs)~\cite{moody20222022} offering a means of miniaturising complex quantum optics experiments that would otherwise be bolted onto heavy vibration-isolation optical benches. By harnessing the innate optical phase stability of PICs, large scale photonic circuits with a high density of components can be used readily \cite{qiang2018large}. This, alongside complementary metal-oxide-semiconductor (CMOS) mass-manufacturing techniques \cite{politi2008silica}, promises a route to scalable photonic quantum technologies.

Quantum states of the electromagnetic field are represented using either a discrete-variable (DV) or continuous-variable (CV) approach. The DV regime defines states using the (discrete) photon number, known as the Fock basis, while in contrast the CV regime defines states using the (continuous) electric field quadratures in a phase-space representation. Both cases lead to a description of quantum states of light that depart entirely from its classical description. For example, Fock states exhibit a photon-number variance which is lower than classical light, while squeezed states are characterised by the lower-than-shot-noise fluctuations of the electric field at certain phases.

The motivation for operating in the CV regime is the ability to generate and entangle states deterministically \cite{yukawa2008experimental,zhuang2018distributed}. Furthermore, CV state detection and characterisation is possible using readily-available room temperature, high-efficiency photodiodes, often operating in a high-bandwidth optical homodyne detection configuration \cite{kumar2012versatile}.

The possibilities of CV quantum technologies
%of the CV domain 
are already being mapped out, and demonstrated, in a number of applications. CV quantum key distribution (QKD) is implementable with classical telecommunications infrastructure \cite{jouguet2014high}. In the field of metrology and sensing, access to a sub-shot noise regime presents unprecedented advantages \cite{giovannetti2006quantum,qin2023unconditional,pradyumna2020twin}. Squeezed light has been used for demonstrations of enhanced biological measurements \cite{taylor2014subdiffraction}, microscopy techniques \cite{atkinson2021quantum} and multi-mode estimation protocols with distributed quantum sensing of phase shifts \cite{guo2020distributed}, that could manifest for example as local changes in refractive index. 
\begin{figure*}[htbp]
    \centering 
    \includegraphics[width=\linewidth]{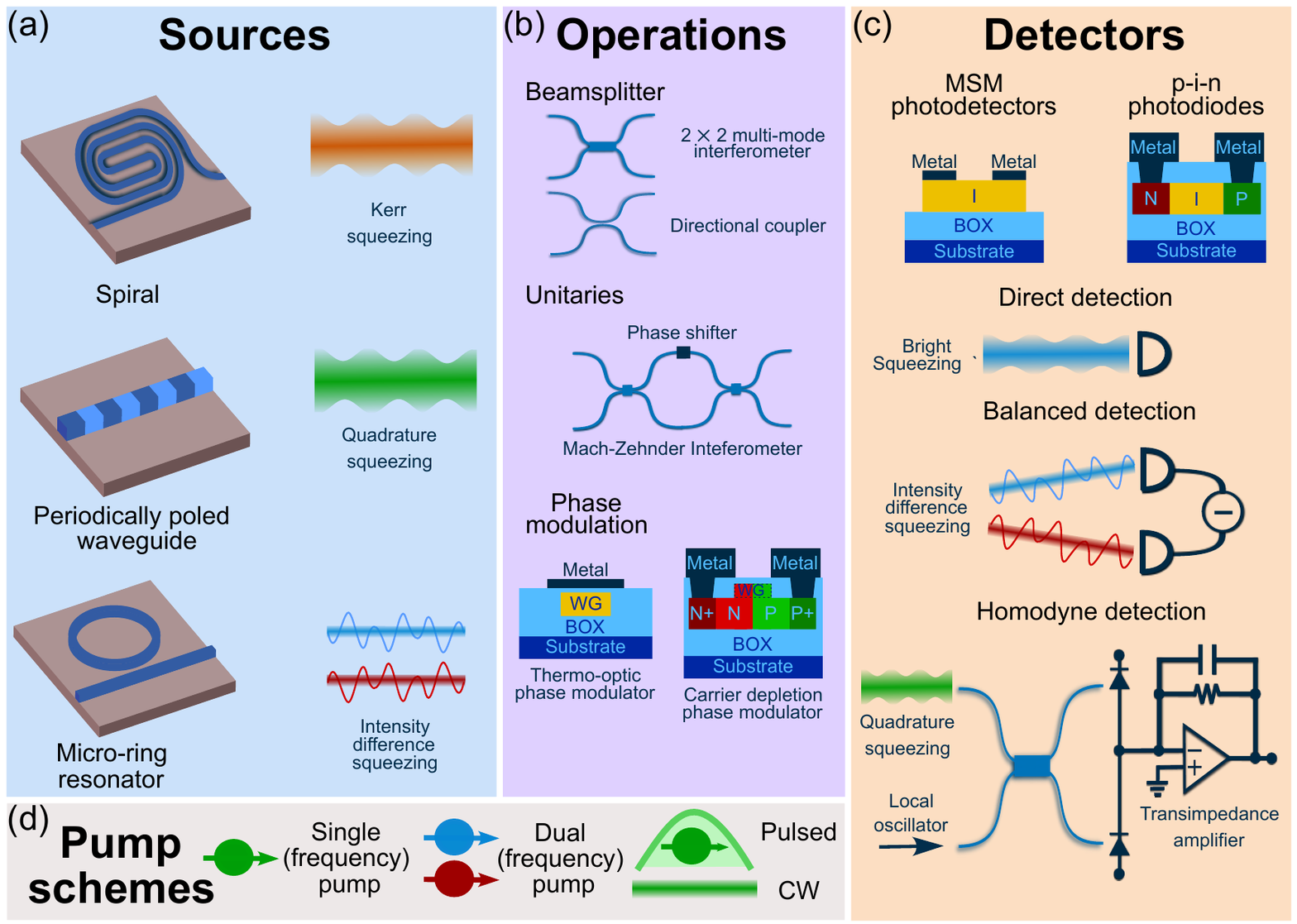}
    \caption*{\textbf{Figure 1}: Components towards a monolithic chip for squeezed light generation, manipulation, detection and readout. (a) Structures for generated squeezed light: spiral waveguides, periodically poled waveguides and micro-ring resonators. Spiral waveguides with a long interaction length are commonly used for mediating self- and cross-phase modulation processes, which can generate a Kerr state. Periodically poled waveguides are engineered to quasi-phase match a nonlinear process that can generate broadband quadrature squeezed light. Micro-ring resonators are commonly used as an optical parametric oscillator (OPO), where below-threshold operation results in quadrature squeezed light while above-threshold operation generates bright squeezed light, observed as for example intensity squeezing between twin beams. (b) Components for performing operations on squeezed light states. Beamsplitter and unitary operations are implemented using multi-mode interferometers, directional couplers, and tunable Mach-Zehnder interferometers (MZI). Phase modulation in the silicon material family relies on thermo-optic phase modulators (TOPM) or carrier depletion phase modulators (CDPM). They can be implemented for tuning device resonances and, when in an MZI, as a switch. TOPMs are simpler to fabricate but limited to $\sim$ \si{\kilo\hertz} for switching operations, while CDPMs are larger structures requiring more complex fabrication but offer $\sim$ \si{\giga\hertz} switching. WG = waveguide, BOX = buried oxide. (c) Photodetector device stacks for MSM and p-i-n photodiodes, as well as detection schemes for different squeezed states. Bright squeezed states, specifically amplitude-squeezed states, can be observed with direct detection on a single photodetector. Intensity difference squeezing between bright twin beams requires balanced detection. All other squeezed states outside of these two cases will require a phase-selective measurement through homodyne detection. P = p-doped semiconductor, N = n-doped semiconductor, I = intrinsic semiconductor. (d) Pump schemes for squeezed light generation, which must be carefully prepared and selected depending on the nonlinear process and desired output squeezing. } 
    \label{fig:Fig1}
\end{figure*}
CV states may be employed in quantum information processing (QIP) -- a thorough description of CV state preparation, operation and detection can be found in a detailed review by Andersen \textit{et al.}~\cite{andersen2010continuous}. Owing to their deterministic generation, squeezed states form the backbone of many CV QIP schemes\cite{lloyd1999quantum,menicucci2006universal,fukui2022building}. With applications across communications, sensing and QIP, the ability to generate and detect high quality squeezed states is integral in realising the potential of CV quantum photonic technologies.

The ultimate goal of monolithic integration of quantum light generation and detection, with the addition of readout and control electronics in an electronic-photonic integrated circuit or `ePIC'~\cite{tasker2024bi} (Figure 1), presents a compelling pathway to smaller footprint, inherently stable and manufacturable quantum technologies. This motivates the development of integrated CV devices, material platforms and fabrication processes. This article will review the experimental developments of integrated continuous-variable systems, from demonstrations of on-chip squeezing, to integrated photodetectors and homodyne detectors, within the CMOS-compatible platforms. 

%%%%%%%%%%%%%%%%%%%%%%%%%%%%%%%%
%BOX 1
\begin{figure*}
%\centering
    \begin{framed}
        \begin{subfigure}[b]{0.47\textwidth}
        \justifying
        \Large
        \textbf{BOX 1: CV state toolkit}
        \vspace{4pt}

        \small
        \textbf{CV states}: The distribution of a CV state in phase space is represented by a quasi-probability distribution, commonly the Wigner function, defined over the orthogonal $P$ and $X$ quadratures with an amplitude $\alpha$ and phase $\varphi$. A state with a mean coherent amplitude much greater than zero, $\vert \bar{\alpha}\vert \gg 0$, is referred to as a `bright' state, with bright squeezed states classified as a specific case of squeezed coherent states; while the state with a mean coherent amplitude of zero, $\vert\bar{\alpha}\vert = 0$ corresponds to the vacuum state. A squeezed vacuum state may be transformed into a bright squeezed state via the displacement operator \cite{lvovsky2015squeezed}. The figure below shows some examples of common optical states: vacuum, coherent and squeezed. These states are referred to as `Gaussian' states as their Wigner function is Gaussian in shape.

        \vspace{6pt}

        \includegraphics[width=0.9\linewidth]{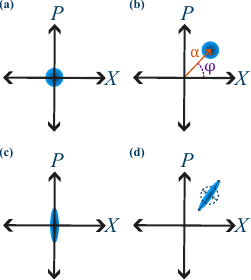}
        \footnotesize
        \textbf{Phase space representations}: (a) The vacuum state; (b) the coherent state; (c) a squeezed vacuum state; (d) a squeezed coherent state. In (b), $\alpha$ denotes the amplitude and $\varphi$ the phase of the state. In (d), the inward arrows denote the squeezed quadrature and the outward arrows denote the antisqueezed quadrature, with the dashed circle indicating the coherent state. 
        
        \vspace{4pt}
    
        \small
        Squeezed states can be generated in two forms. If the generated squeezed state occupies a single optical mode as defined by its spatio-temporal profile, frequency, and polarisation, and it can be described by the operators $\hat{a}_{1}$ and $\hat{a}_{1}^{\dagger}$, we can define the single-mode squeezed state (SMSS) Hamiltonian as 
         \begin{equation}
        \hat{H}_\mathrm{SM} = i\hbar\alpha\left[\hat{a}_{1}^{2} - \left({\hat{a}_{1}^{\dagger}}\right)^{2}\right]/2.
        \label{eq:SMSqueezing}
        \end{equation}
        If the generated squeezed state instead occupies two optical modes described by the operators $\hat{a}_{1}$, $\hat{a}_{1}^{\dagger}$ and $\hat{a}_{2}$, $\hat{a}_{2}^{\dagger}$ respectively, the two-mode squeezed state (TMSS) Hamiltonian is given by
        \begin{equation}
         \hat{H}_\mathrm{TM} = i\hbar\alpha\left[\hat{a}_{1}\hat{a}_{2} - \hat{a}_{1}^{\dagger}\hat{a}_{2}^{\dagger}\right]/2.
        \label{eq:TMSqueezing}
        \end{equation}

        The effect of linear losses on a squeezed state can be modelled with a beamsplitter transformation. The quadrature variance of the signal, $\langle\Delta\left( X_{a,\theta} \right)^{2}\rangle$ is transformed by the effective beamsplitter transmission $T$ into the measured quadrature variance $\langle\Delta\left( X_{\theta,a'} \right)^{2}\rangle$ by Equation \ref{eq:SqueezingLoss}
        \end{subfigure}
        \begin{subfigure}[b]{0.47\textwidth}
        \justifying
        
        \begin{equation}
         \langle\Delta\left( X_{\theta,a'} \right)^{2}\rangle = T\langle\Delta\left( X_{a,\theta} \right)^{2}\rangle + \left(1-T\right)/2 .
        \label{eq:SqueezingLoss}
        \end{equation}
        This relation demonstrates how experimental losses can be estimated by measuring the deviation of the quadrature variance from the minimum uncertainty. Furthermore, linear loss cannot solely obscure finite squeezing; even at very low $T$ values there will remain some observable squeezing in the measured quadrature variance $\langle\Delta\left( X_{\theta,a'} \right)^{2}\rangle$.
        
        \vspace{0.2cm}
        
        \textbf{Nonlinear processes}: The use of nonlinear optics to generate single- and two-mode squeezing is well-researched \cite{andersen201630}. A third-order parametric nonlinear process, requiring a medium with a sufficiently large $\chi^{(3)}$ component, involves four separate light fields interacting in a way that conserves optical energy.  The process of two pump photons creating a signal and idler pair is known as four-wave mixing (FWM). The Hamiltonian governing this process is given by 
        \begin{equation}
            \hat{H}_{\mathrm{FWM}} = -\hbar g_{0} \left[ \hat{a}_{p1}^{\dagger}\hat{a}_{p2}^{\dagger}\hat{a}_{s}\hat{a}_{i} - \hat{a}_{p1}\hat{a}_{p2}\hat{a}_{s}^{\dagger}\hat{a}_{i}^{\dagger}\right].
            \label{eq:H_FWM}
        \end{equation}
        In the case of a strong pump field, the limit where $\hat{a}_{p1,2}\rightarrow A_{p1,2}$, Equation \ref{eq:H_FWM} can be rewritten as
        \begin{equation}
            \hat{H}_{\mathrm{FWM'}} = -\hbar g_{0} \left[A_{p1}A_{p2} a_{s}a_{i} - A_{p1}^{*}A_{p2}^{*}a_{s}^{\dagger}a_{i}^{\dagger}\right].
            \label{eq:H_FWM_p}
        \end{equation}
        In the degenerate case of equal signal and idler frequencies, Equation \ref{eq:H_FWM_p} has the same form as Equation \ref{eq:SMSqueezing} and a SMSS will be generated; conversely the non-degenerate case has the same form as Equation \ref{eq:TMSqueezing} and thus a TMSS is generated. We note that the second-order parametric nonlinear process of spontaneous parametric down conversion is also able to produce SMSS and TMSS by converting an incoming pump into outgoing signal and idler fields that share the same frequency mode.

        In accordance with conservation of energy, certain wave-mixing processes can provide parametric gain for incoming optical modes. This is known as optical parametric amplification (OPA), and can be stimulated with a continuous energy transfer from the pump to the signal along the nonlinear medium length \cite{Agrawal2011}. 
        A high-gain OPA can noiselessly amplify the in-phase quadrature of an incoming state, while the orthogonal quadrature is deamplified. If the incoming state is vacuum, this gives rise to quadrature squeezing; if the incoming state is coherent, this gives rise to bright squeezing. For any general input state, this phase-sensitive amplification may be used for noise-tolerant readout of CV states \cite{breitenbach1997measurement}. Parametric amplification in a cavity is known as an optical parametric oscillator (OPO). The weak outgoing fields of nonlinear processes are enhanced in an OPO due to both the parametric gain and the cavity itself. The cavity threshold dictates the resulting squeezed state: pumping below the threshold will result in a single-mode or two-mode squeezed vacuum state (SMSS, TMSS), while pumping above threshold will result in bright intensity-difference squeezing between the signal and idler beams \cite{fabre1989noise}. In addition, the OPO sets the maximum observed bandwidth of the squeezing through the cavity dynamics.
        
        \vspace{0.2cm}
        
        The optical Kerr effect, resulting from the intensity-dependent refractive index term for $\chi^{(3)}$ materials, leads to self-phase modulation (SPM) and cross-phase modulation (XPM). A strong incident pump field in a nonlinear medium will couple to itself via the Kerr interaction, giving rise to a nonlinear refractive index and further, an intensity-dependent phase evolution. This has the overall effect of the field acquiring a nonlinear phase shift as it propagates through the medium \cite{sirleto2023introduction}.

        \end{subfigure}
    \end{framed}
\end{figure*}

\begin{figure*}
%\centering
    \begin{framed}
        \begin{subfigure}[b]{0.48\textwidth}
        \justifying
        This process stretches out the circular coherent state in phase space into an elliptical-shaped distribution. The figure below shows an illustration of how bright squeezed states can emerge from the Kerr effect. Whilst the Hamiltonian of this intensity-dependent phase shift does not match that of the operator for quadrature squeezing, the shape of its phase space distribution in the low Kerr limit approximates a displaced squeezed state.  
        
        \includegraphics[width=0.8\linewidth]{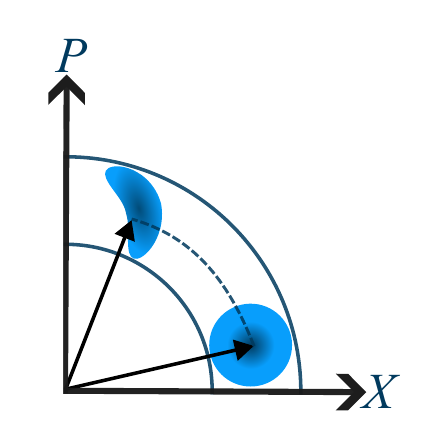}
        \footnotesize
        
        \textbf{The Kerr state}: a coherent state propagating through a Kerr medium will acquire a nonlinear phase shift, resulting in the ``crescent'' shape in phase space.
        
        \vspace{4pt}
    
        \small
        This crescent-shaped ellipse, where certain quadrature components are squeezed, preserves the photon number and  results in a bright squeezed state. Unlike FWM, which requires careful device engineering for phase-matching, squeezing produced by the Kerr interaction is inherently phase matched. However a pulsed pump is typically used which requires careful control of the pulse profile.

        \vspace{0.2cm}

        \textbf{Materials for integrated devices}: To produce squeezed light states, a parametric process and therefore materials with a nonlinear susceptibility component ($\chi^{(2)}$ or $\chi^{(3)}$) are required. Discussion in this review will focus on $\chi^{(3)}$ materials for  generating squeezed light on-chip.  
        \end{subfigure}
        \begin{subfigure}[b]{0.48\textwidth}
        \justifying
        \small
        
        The ability to confine and guide light on the nanoscale presents advantages for both scalable manufacturing, and accessing a regime of intense fields by reducing the mode volume. A natural first choice for devices to attempt to scale-up has been the CMOS-compatible family, namely silicon (Si), silicon nitride (SiN), silicon dioxide (SiO\textsubscript{2}) and germanium (Ge), which are able to leverage a mature fabrication infrastructure. Additionally, high refractive index contrast between the waveguiding layer and surrounding cladding permits good mode confinement, reducing losses and enabling a reduced device footprint. These platforms will be the focus of this review, however the authors note that there is also interest in a wider range of material systems owing to a variety of desirable optoelectronic properties, including a signfiicant $\chi^{(2)}$ component. Examples include lithium niobate (LiNbO\textsubscript{3}), barium titanate (BTO), and III-V semiconductor systems including aluminium nitride (AlN),  gallium arsenide (GaAs) and indium phosphide (InP), which have all been successfully integrated using thin film and epitaxial techniques. A comprehensive description of nonlinear optical materials for integrated applications can be found in the recent review by Dutt \textit{et al.} \cite{dutt2024nonlinear}.

        Materials that host nonlinear processes are also subject to non-parametric processes which manifest as nonlinear losses, transferring energy from the mixing process to the material, for example via optical phonons or electron levels \cite{sirleto2023introduction}. Additionally, the benefits of integration become a trade-off as sensitivity to these losses increases. Two photon absorption (TPA) occurs for high optical intensities, where two photons with combined energies spanning the material band gap may be absorbed, promoting a charge carrier in the material \cite{leuthold2010nonlinear}. This may lead to free carrier absorption (FCA), where unbounded charge carriers continue to exacerbate the losses. These two mechanisms are particularly deleterious in narrow band-gap semiconductors. Other non-parametric losses include scattering processes, such as Raman or Brillouin scattering, resulting in photons being redirected away from the desired optical mode. Parametric processes themselves may be detrimental if they siphon photons away from or add noise into the desired signal mode - for example, Bragg-scattering FWM.
        \end{subfigure}
    \end{framed}
\end{figure*}

\section{Integrated sources}
The generation of squeezing on chip is possible using nonlinear materials and the general toolkit provided by integrated photonics (Box 1). An optical parametric amplifier (OPA) can be formed with an integrated waveguide while cavity structures such as microring resonators are the basis for an optical parametric oscillator (OPO). Silicon (Si) and silicon nitride (Si\textsubscript{3}N\textsubscript{4}) have a sufficiently high $\chi^{(3)}$ component to mediate third-order nonlinear processes. Si exhibits a higher $\chi^{(3)}$ nonlinearity and refractive index, allowing smaller footprint devices due to better mode confinement. Furthermore its maturity and ubiquity in the microelectronics industry have driven its use as the principal platform of choice for integrated photonics. However, its use in nonlinear applications is limited by the undesirable two-photon absorption (TPA) and free-carrier absorption (FCA) at telecom wavelengths \cite{leuthold2010nonlinear}. Si\textsubscript{3}N\textsubscript{4} has a reduced sensitivity to TPA and FCA at telecom wavelengths, and devices have been reported across a broad wavelength range \cite{yang2024visible} with very low losses \cite{bose2024anneal}. % waveguides \cite{bose2024anneal}, high quality (Q)-factor devices \cite{luke2013overcoming} and  OPOs \cite{levy2010cmos}. 
The additional benefits of Si\textsubscript{3}N\textsubscript{4} is the ease of deposition using techniques such as plasma-enhanced chemical vapour deposition (PECVD), a CMOS-compatible process, and low-pressure chemical vapour deposition (LPCVD), an approach that has resulted in very low-loss films \cite{dutt2024nonlinear}.  However, Si\textsubscript{3}N\textsubscript{4} resonator devices suffer from thermorefractive phase noise \cite{huang2019thermorefractive}, where the material refractive index can change with temperature fluctuations via the thermo-optic effect.

\subsection{Integrated squeezers}
\label{ssec:IntegratedSqueezers}

Squeezing generated on an integrated Si\textsubscript{3}N\textsubscript{4} chip was first reported by Dutt \textit{et al.} \cite{dutt2015chip}, measuring \SI{1.7}{\decibel} of intensity-difference squeezing in bright twin beams over a \SI{5}{\mega\hertz} bandwidth. The device comprised Si\textsubscript{3}N\textsubscript{4} microring resonators \cite{levy2010cmos,luke2013overcoming}, forming an OPO able to mediate a four-wave mixing (FWM) process. Correcting for detection and off-coupling losses resulted in an estimated \SI{5}{\decibel} of generated squeezing, which was assumed to be limited by excess pump noise given that the device was operating above-threshold. The same group demonstrated the tunability of the squeezing level by adjusting the coupling condition of the microring resonator \cite{dutt2016tunable}. K\"ogler \textit{et al.} \cite{alfredo2024quantum} performed quantum state tomography of the four-mode state, comprising the upper and lower sidebands of signal and idler twin beams, produced from a Si\textsubscript{3}N\textsubscript{4} microring resonator OPO pumped above threshold. Using a similar experimental set up to that in \cite{dutt2015chip}, \SI{2.3}{\decibel} of intensity-difference squeezing was measured, inferred as \SI{4.9}{\decibel} on-chip. The discrepancy between the inferred squeezing level and the estimate of \SI{9}{\decibel} for this OPO was attributed to Kerr-effect phase modulations distorting the noise ellipse, which worsened with pump power. Very recently, a group led by Dutt demonstrated \SI{3.7}{\decibel} of directly measured intensity-difference squeezing from Si\textsubscript{3}N\textsubscript{4} microring resonators, with \SI{10.2}{\decibel} inferred on chip \cite{shen2025strong}. The measured level of squeezing is in agreement with calculations based on the coupling coefficient and detection losses, demonstrating that excess classical noise from the pump has been sufficiently suppressed. This was achieved by using a microring with a much higher free spectral range (\SI{80}{\giga\hertz} in \cite{dutt2015chip}, \SI{450}{\giga\hertz} in \cite{shen2025strong}), which lowers the OPO threshold. This in turn reduces the intracavity power, improves the OPO stability and ensures that any classical noise can be suppressed with the balanced detection.

The 2020 work by Cernansky \textit{et al.} \cite{cernansky2020nanophotonic} demonstrates a single-mode squeezed state (SMSS) generated via counter-propagating bright squeezed states in a Sagnac interferometer with Si\textsubscript{3}N\textsubscript{4} microring resonators. Light enters the Sagnac loop via a multimode interferometer (MMI), engineered to split TE light 50:50, and TM light 59:41. The TE light splits into two counter propagating beams which undergo Kerr squeezing due to self-phase modulation \cite{hoff2015integrated} within the ring tuned onto resonance with the TE field. These two bright beams reinterfere at the Sagnac output, effectively cancelling out the coherent amplitude to result in a quadrature squeezed state rather than a displaced squeezed state. Simultaneously, TM light is transmitted unequally and co-propagates with the bright TE beams while being detuned from the ring resonance, partially interfering at the output, to be used as a phase-stable LO. A maximum of \SI{0.45}{\decibel} squeezing over a bandwidth of \SI{300}{\mega\hertz} is measured, with an inferred squeezing of \SI{1}{\decibel} on-chip. The squeezing is limited by thermorefractive noise, which is characterised in the supplementary materials; the authors conclude that the classically correlated thermorefractive noise transferred through the Sagnac loop could be reduced with a higher interferometer contrast.

Vaidya \textit{et al.} \cite{vaidya2020broadband} report both quadrature squeezing and photon number difference squeezing in Si\textsubscript{3}N\textsubscript{4} microring devices. The states generated were shown to be compatiable with requirements for quantum sampling applications \cite{vernon2019scalable}. A maximum of \SI{1}{\decibel} for two-mode squeezed vacuum was measured, estimated as \SI{4}{\decibel} on-chip when correcting for losses, observed over a \SI{1}{\giga\hertz} bandwidth. Photon number difference correlations were also measured on a higher Q-factor device, allowing a weaker pump signal and improving the signal-to-noise ratio. For this measurement, signal and idler pairs were separated with wavelength division multiplexing (WDM) components and incident upon photon number resolving detectors. At the highest pump powers, corresponding to the optimal signal-to-noise ratios, the highest measured photon number difference squeezing is \SI{1.5}{\decibel}, inferred to be \SI{7}{\decibel} on chip due to the finite escape efficiency of the resonator. The limiting factor for both devices is considered to be out-coupling and propagation losses. A later Si\textsubscript{3}N\textsubscript{4} device from Xanadu, based on coupled ring resonators of different sizes - accordingly named the `nanophotonic molecule' \cite{zhang2021squeezed} - demonstrated up to \SI{1.65}{\decibel} of SMSS measured over \SI{1}{\giga\hertz}. In this device, two rings referred to as the `auxiliary' and `principal', were fabricated with different geometries and Q-factors to generate a set of hybridised modes. This technique allowed the suppression of parasitic processes that siphon photons away from the desired dual-pump FWM mode, including Bragg-scattered and single-pump FWM photons.

The desire for a large number of squeezed modes through frequency multiplexing has encouraged demonstrations of squeezed microcombs. Yang \textit{et al.} demonstrated up to twenty qumode pairs of TMSS over a frequency comb spanning \SI{1}{\tera\hertz}, from a silica wedge microresonator device \cite{yang2021squeezed}. By pumping the microring with a single resonant pump, multimode squeezed vacuum is generated over many nearby resonances akin to a frequency comb. Jahanbozorgi \textit{et al.} \cite{jahanbozorgi2023generation} employed a Si\textsubscript{3}N\textsubscript{4}  microresonator to generate seventy qumode pairs of TMSS over \SI{1.3}{\tera\hertz}. A maximum of \SI{1.1}{\decibel} quadrature squeezing was directly measured, with \SI{4.4}{\decibel} inferred. Recent work by Jia \textit{et al.} \cite{jia2025continuous} demonstrate a Si\textsubscript{3}N\textsubscript{4}  microresonator pumped by a phase-locked optical frequency comb to generate supermode-based multipartite entangled states, with a maximum squeezing of \SI{2}{\decibel} observed.

\begin{figure*}[htbp]
    \centering 
    \includegraphics[width=\linewidth]{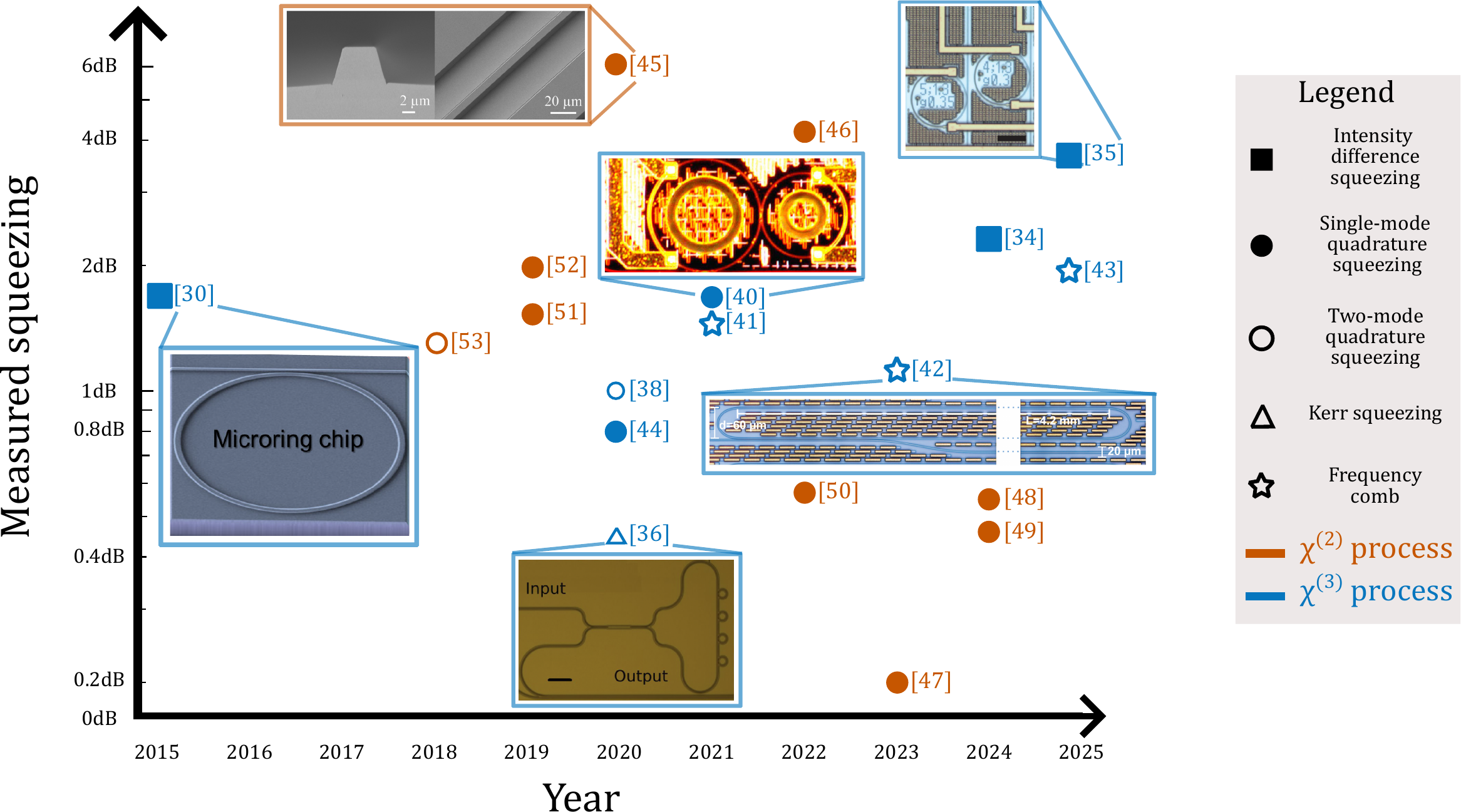}
    \caption*{\textbf{Figure 2}: Integrated devices that have generated squeezed light states using different nonlinear optical processes over the last decade, where the $y$-axis is the measured squeezing for each device. Devices are coloured to indicate the nonlinearity of the material, and the shapes refer to the generated squeezing. The $\chi^{\left( 3 \right)}$ devices (blue) use the Si\textsubscript{3}N\textsubscript{4} platform (except reference \cite{yang2021squeezed} which is SiO\textsubscript{2}), while the $\chi^{\left( 2 \right)}$ devices (orange) use the LiNbO\textsubscript{3} platform.  In references \cite{dutt2015chip,alfredo2024quantum,shen2025strong}, a Si\textsubscript{3}N\textsubscript{4} microring resonator was used as an OPO and operated above threshold to produce intensity-difference squeezing, while references \cite{vaidya2020broadband,zhang2021squeezed,zhao2020near} used Si\textsubscript{3}N\textsubscript{4} microring resonators with different pump configurations below threshold to produce single- or two-mode squeezed vacuum. Cernansky \textit{et. al.} \cite{cernansky2020nanophotonic} generate squeezed vacuum by interfering bright Kerr states in a Sagnac interferometer. Works in \cite{yang2021squeezed,jahanbozorgi2023generation,jia2025continuous} use Si\textsubscript{3}N\textsubscript{4} microring resonators as frequency comb sources of squeezed light. LiNbO\textsubscript{3} devices are implemented as an OPA in works \cite{kashiwazaki2020continuous,nehra2022few,stokowski2023integrated} and an OPO in \cite{park2024single,arge2024demonstration}. LiNbO\textsubscript{3} waveguides \cite{chen2022ultra}, resonators \cite{otterpohl2019squeezed} and PICs are implemented to mediate SPDC \cite{mondain2019chip} and entangled two-mode states \cite{lenzini2018integrated}. } 
    \label{fig:Fig2}
\end{figure*}

We briefly note that there have been many examples of integrated squeezing in LiNbO\textsubscript{3}, which has a large $\chi^{(2)}$ component, using integrated waveguides for spontaneous parametric down conversion (SPDC) and optical parametric amplification (OPA). Engineering techniques such as periodic poling for phase-matching have resulted in broadband and high levels of squeezing. Kashiwazaki \textit{et al.} \cite{kashiwazaki2020continuous} used a single-pass configuration with a periodically poled lithium niobate (PPLN) waveguide to measure \SI{4.5}{\decibel} of squeezing over \SI{250}{\mega\hertz}, with a total estimated bandwidth of \SI{2.5}{\tera\hertz}. Chen \textit{et. al.} \cite{chen2022ultra} measured \SI{0.56}{\decibel} of squeezing at $\sim$\SI{1550}{\nano\metre} over a \SI{7}{\tera\hertz} bandwidth with a double-PPLN waveguide configuration to perform SHG followed by SPDC. Nehra \textit{et. al.} \cite{nehra2022few} performed parametric amplification of a vacuum-squeezed state, measuring \SI{4.2}{\decibel} of squeezing at \SI{2090}{\nano\metre} over a \SI{25.1}{\tera\hertz} bandwidth, inferring a total of \SI{10.4}{\decibel} of squeezing available over \SI{36.4}{\tera\hertz}. Other works have integrated LiNbO\textsubscript{3} squeezers into PICs \cite{mondain2019chip,stokowski2023integrated,park2024single}, demonstrated reconfigurable deterministic entanglement generation \cite{lenzini2018integrated}, and even generated squeezing without periodic poling \cite{arge2024demonstration}. Figure 2 summarises the key device parameters and references of integrated sources of squeezed light states.

\subsection{Coherent sources}
\label{ssec:coherent}
Some implementations of CV quantum technologies require a coherent rather than a squeezed state. Early integrated lasers were predominantly based on III-V semiconductors due to their direct bandgap. Ongoing developments of lasers on Si typically rely on heterogeneous integration, for example wafer bonding or transfer printing \cite{bao2021review}. Advances within the Si\textsubscript{3}N\textsubscript{4} platform include microring resonators for parametric oscillation across the visible \cite{corato2023widely} and into the infrared \cite{levy2010cmos}. For further details, we direct the reader to recent reviews on integrated laser technologies \cite{li2022integrated,lu2024emerging}.

\begin{figure*}
%\centering
    \begin{framed}
        \begin{subfigure}[b]{0.48\textwidth}
        \justifying
        \Large
        \textbf{BOX 2: CV detection schemes}
        
        \vspace{4pt}
        
        \small
        \textbf{Detecting squeezed light}: The type of measurement required to probe squeezed light will depend on the generated state, however a suitable choice of photodetector is crucial for successful detection. Bright squeezed states can be observed with direct detection on a single photodetector, intensity-difference squeezing across bright twin beams can be measured with balanced detection over two photodetectors, and quadrature squeezing requires optical homodyne detection \cite{andersen2010continuous}. Figure 1 includes schematics for each of these methods. For all methods, photodetectors with high quantum efficiencies and fast response times are desirable: high efficiency is required as the quantum signals are weak and vulnerable to loss, while response times may limit the end application of the device.

        Photodetectors are characterised with these key performance metrics. The responsivity relates the detector quantum efficiency to its spectral sensitivity, while the dark current governs the detector’s noise floor. Both parameters depend on the material used in the absorbing region of the detector, particularly the bandgap properties, absorption coefficient and material quality. The detector bandwidth ultimately limits the resolvable signal speed, and is related to the charge carrier transit times which typically depend on device material and geometry. % \cite{sze2021physics}.
        
        \vspace{0.5cm}
        
        \textbf{Homodyne detection}: Homodyne detection is an experimental technique applied to oscillating signals to obtain information contained within the signal modulation. In quantum optics, this Gaussian measurement is typically used for the measurement and tomography of CV states of light, consisting of a 50:50 beam splitter on which the weak quantum signal is interfered with a bright, classical local oscillator (LO). The two outputs of the beam splitter are incident on two photodetectors, and by subtracting the photocurrents such that the classical noise is cancelled out, information can be obtained about the quantum signal's quadrature at a phase specified by the relative phase between the quantum state and the LO. At the output of the homodyne detector, the subtracted photocurrent is weak and requires electronic amplification for practical applications. Therefore the design of an optical homodyne detector also involves the design of an electronic transimpedance amplifier (TIA) circuit. In order to measure the weak quantum fields, the amplification and readout electronics need to have sufficiently low noise to optimise the sensitivity of the homodyne detector. This is typically characterised by the shot noise clearance (SNC), which defines how far the total measured signal noise is above the detector noise. In combination with the photodetector responsivity, the SNC defines the homodyne detector efficiency. The common mode rejection ratio (CMRR) quantifies how well the balancing of the homodyne detector cancels out the LO classical noise common to both photodetectors \cite{Jin2015}. While also sufficiently rejecting strong classical noise that can obscure the weak quantum signal of interest, a high CMRR prevents a DC current corresponding to the LO from flowing into the TIA and reducing its dynamic range \cite{bruynsteen2021integrated}. The CMRR of integrated homodyne detectors benefit from inherent mode matching in the waveguide structures and the high precision for path length matching \cite{bruynsteen2021integrated}. An MZI structure with a phase shifter before detection allows for precise tuning of the balancing, compensating for fabrication intolerances of the paths or detectors \cite{tasker2021silicon}. In CV quantum computing and CV-QKD, the detection bandwidth impacts the clock rate and secret-key rate respectively \cite{Asavanat2021,Liu2015}. Thus, optimising the detection bandwidth is a key factor for  
        \end{subfigure}
        \begin{subfigure}[b]{0.48\textwidth}
        \justifying

        \includegraphics[width=0.8\textwidth]{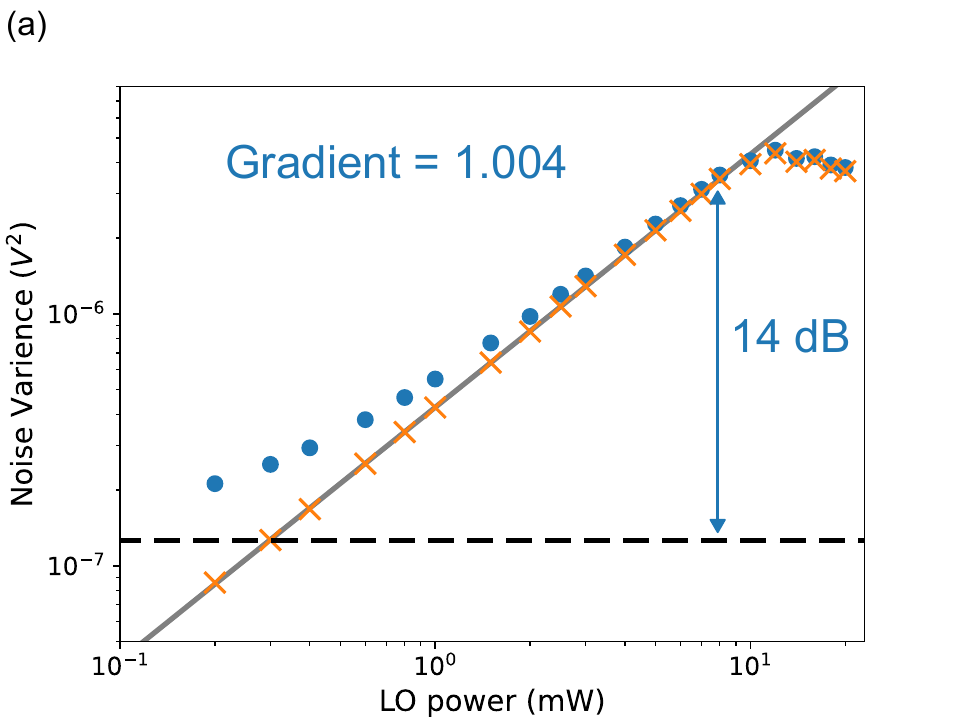}

        \includegraphics[width=0.8\textwidth]{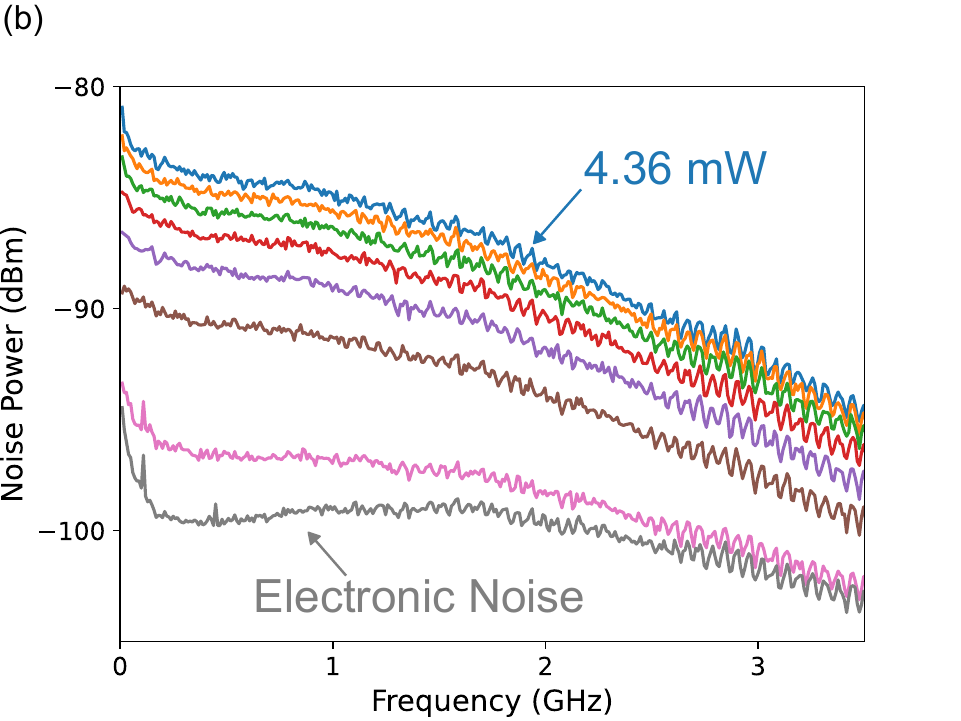}

        \includegraphics[width=0.8\textwidth]{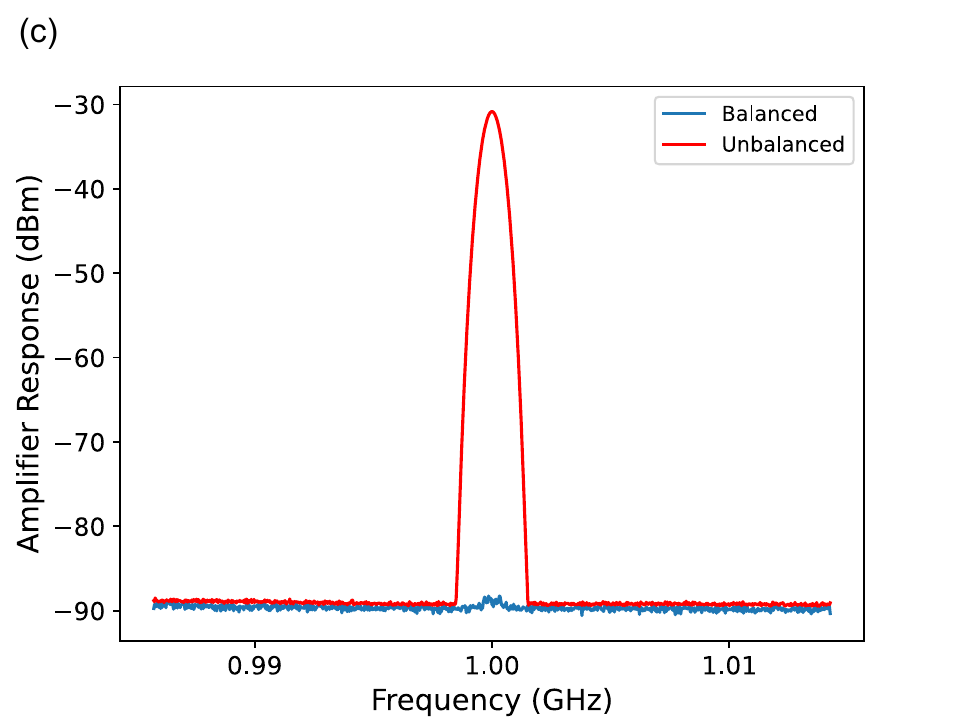}

        \footnotesize
        
        \textbf{Homodyne detector characteristics}: (a) The SNC, (b) bandwidth response with increasing LO power and, (c) CMRR of a homodyne detector. Figures adapted from Tasker \textit{et al.} \cite{tasker2021silicon}.
        
        \vspace{4pt}
    
        \small
        scalable CV technologies \cite{bruynsteen2021integrated,tasker2021silicon,tasker2024bi,Ng2024}, typically achieved with the appropriate choice of photodetectors and TIA. With fast integrated photodiodes available, the TIA circuits are currently the limiting factor of homodyne detection bandwidth. To obtain amplification that is flat in frequency, homodyne detector TIAs are designed to have a Butterworth response with a 3dB bandwidth given by

        \begin{equation}
            f_{3\mathrm{dB}}=\sqrt{\frac{A_{0}f_{0}}{2\pi R_{F}C_{\mathrm{tot}}}},
        \label{eq:Butterworth}
        \end{equation}

        where $A_{0}f_{0}$ is the gain-bandwidth product of the operational amplifier (op-amp), $R_{F}$ is the feedback resistor of the TIA circuit and $C_{\mathrm{tot}}$ is the total capacitance. The total capacitance consists of the inherent capacitance of the photodiodes and the op-amp, the feedback capacitance used to flatten the TIA frequency spectrum as well as the parasitic contributions \cite{Masalov2017}. For integrated homodyne detectors wire-bonded to a discrete TIA circuit on a PCB, the parasitic capacitance of the PCB dominates $C_{\mathrm{tot}}$, limiting the detector bandwidth \cite{raffaelli2018homodyne,tasker2021silicon}. This motivates the full integration of the PIC with the electronics into a monolithic ePIC.

        \end{subfigure}
    \end{framed}
\end{figure*}

\section{Integrated detectors}
\label{sec:IntegratedDetectors}
The development of high-bandwidth detectors that are CMOS-compatible for the integration of PICs is of interest for both classical information processing \cite{Benedikovic2019,Peng2024} and quantum photonics \cite{tasker2021silicon,bruynsteen2021integrated}. Detection techniques vary according to the requirement set by the CV state to be measured (Box 2). Photodetectors are typically based on two principal paradigms: metal-semiconductor-metal (MSM) structures, where optically created charge carriers are separated by an externally applied electric field (bias); and active diode structures that use dopants to create an internal field for charge carrier separation (p-i-n diodes and pn-junctions).

\subsection{Integrated photodetectors}
\label{sec:IntegratedPDs}

Metal-semiconductor metal (MSM) detectors are simple devices to fabricate on-chip, requiring only a pair of metallic contacts placed over the semiconductor region.  The detector performance is controlled by geometry: silicon has a relatively long absorption depth (compared to wavelength) for sub 1\,\textmu m light, high efficiency detectors thus need to be tens of \si{\micro\metre} long; 70\,\textmu m long detectors have been reported to reach external quantum efficiencies of 60\%  \cite{Ma2013}.  The record bandwidth performance for such a device is 140\,GHz \cite{Liu1994}: shown in a 5\,\textmu m square detector, 100\,nm thick, the responsivity was exceptionally low (6.7\,mA\,W$^{-1}$) due to low absorption.

Semiconductor doping can create an internal field in the detector structure to separate charges.  p-n junctions have a small depletion region, limiting efficiency; whereas p-i-n diodes create an artificially enlarged depletion region with a central intrinsic layer.  The highest bandwidth PIC detectors are reported in the telecom bands, where the operation of Si-based devices is extended using germanium (Ge): either on top of an existing device (p-i-n) \cite{Xue2024} or embedded within it (p-n) \cite{Benedikovic2019}. Vertically illuminated p-i-n photodiodes have a wide intrinsic region, however the bandwidth is limited by longer transit times\cite{Zebda1994}. Distributed absorption in the wider intrinsic region, introduces excess loss at higher frequencies, becoming signficant above \SI{100}{\mega\hertz} and \SI{1}{\giga\hertz} for Si and InGaAs p-i-n photodiodes respectively. Waveguide-coupled Ge photodiodes break the bandwidth-efficiency trade-off with edge or evanescent coupling between the carrier-collection path and the photon-absorption path, as opposed to free-space illumination \cite{Liu2007}. A recent demonstration uses a narrow region of Ge, within a p-i-n diode, atop the main Si waveguide; and reports bandwidths of 265\,GHz, albeit with a depressed responsivity (\SI{0.3}{\ampere\per\watt}) \cite{lischke2021ultra}.  Additionally, the integration of Ge photodiodes with SOI components is available at many commercial foundries \cite{fedeli2013}.

For shorter wavelengths, all-silicon devices are possible. p-n detectors operating at 685\,nm have reported bandwidths of 19.1\,GHz but at the expense of high dark currents \cite{Yanikgonul2021}; whereas the highest p-i-n device reported is a diffusion-doped detector at 850\,nm with a bandwidth of 14\,GHz and a low dark current of 75\,nA \cite{Chatterjee2019}.  One fabrication challenge with these devices is the semiconductor doping, because ion implantation increases fabrication complexity and can introduce additional propagation losses.  A solution that has been shown is to create a hybrid device: a p-i-n device is fabricated via a commercial multi-project wafer programme, and then transferred via polydimethylsiloxane (PDMS) stamp onto the pre-structured PIC \cite{Cuyvers2022}.

Similarly to squeezing devices based on OPOs (Box 1), microring resonators can provide cavity enhancement of the light-field for enhanced absorption, at the cost of the resonator-limited bandwidth. MSM detectors around a resonator rib demonstrated an enhanced responsivity (\SI{0.81}{\ampere\per\watt} reported \cite{Chatterjee2022}). Peng \textit{et al.} demonstrated a work-around for the limiting bandwidth using a two-stage set of coupled resonators for a p-i-n detector device, with a \SI{40}{\giga\hertz} bandwidth reported.

The detector electrodes limit its performance in 2 ways.  Firstly, metallic surfaces can interact with the optical field introducing plasmonic losses, necessitating some minimum semiconductor width from the optical waveguide channel \cite{Chatterjee2022}.  Secondly, differences in the propagation speed between the optical wave in the detector waveguide, and the electrical wave in the metallic electrode, will smear-out sharply defined detection signals: ultimately limiting the bandwidth.  To overcome this, the system needs to be considered as a transmission line, where the propagation of the electric wave is controlled, as in the demonstration of 140\,GHz bandwidth \cite{Liu1994}, where metallic fingers of controlled impedance contact the Si detector along the propagation direction.

Overall, detector technology for classical telecommunications has been leading a thrust for CMOS-compatible, or all-Si, detector structures in recent years \cite{Yang2019}: providing a wealth of options for integrating detectors directly into PICs for quantum applications.  Specifically, MSM detectors show promise as simple detector structures that can be integrated with low demands on the fabrication workflow.

\subsection{Integrated homodyne detectors}
\label{sec:integratedHDs}

The first demonstration of a PIC-integrated homodyne detector was in 2018, utilised for the tomography of a coherent state and as a quantum random number generator (QRNG) \cite{raffaelli2018homodyne}. In this work the discrete transimpedance amplification (TIA) circuit was implemented on a PCB, onto which the PIC was connected with wirebonds. While the individual photodiode bandwidths were reported as 23 GHz, the bandwidth of the homodyne detector configuration was limited to only 150 MHz by the parasitic capacitances of the discrete electronics packaging and PCB. 
\begin{table*}[htbp]
\centering
 \begin{tabular}{||c | c | c | c | c | c | c ||} 
 \hline 
 \multicolumn{7}{|c|}{\textbf{Integrated photodetectors}} \\
 \hline\hline
 Year & Ref. & Type & $\lambda$ & Bandwidth & Dark Current & Responsivity  \\ 
 \hline\hline
 1994 & \cite{Liu1994} & MSM (Si) & \SI{780}{\nano\metre} & \SI{140}{\giga\hertz} &\SI{0.2}{\pico\ampere\per\centi\metre\squared} & \SI{6.7}{\milli\ampere\per\watt} \\
 1995 & \cite{lee1995novel} & MSM (Si) & \SI{830}{\nano\metre} & \SI{3}{\giga\hertz} & \SI{91}{\nano\ampere} & \SI{0.17}{\ampere\per\watt}  \\
 1999 & \cite{seto1999low} & MSM (Si) & \SI{780}{\nano\metre} & \SI{460}{\mega\hertz} & \SI{1}{\nano\ampere} & \SI{0.02}{\ampere\per\watt}  \\
 2009 & \cite{chen2009ultra} & MSM (Ge) & \SI{1550}{\nano\metre} & \SI{50}{\giga\hertz} & \SI{4}{\micro\ampere} & \SI{0.3}{\ampere\per\watt} \\
 2013 & \cite{Ma2013} & MSM (Si) & \SI{850}{\nano\metre} & -- & 2.5-2.8\si{\micro\ampere} & \SI{0.41}{\ampere\per\watt} \\
 2015 & \cite{li2015over} & p-i-n (Si) & \SI{850}{\nano\metre} & \SI{13.6}{\giga\hertz} & \SI{10}{\pico\ampere} & \SI{7.5}{\milli\ampere\per\watt}  \\ 
 2015 & \cite{Kroh2015} & p-i-n (Ge) & \SI{1550}{\nano\metre} & \SI{70}{\giga\hertz} & \SI{100}{\nano\ampere} & \SI{1}{\ampere\per\watt} \\
 2018 & \cite{sakib2018demonstration} & p-i-n (Si) & \SI{1310}{\nano\metre} & \SI{15.3}{\giga\hertz} & \SI{850}{\nano\ampere} & \SI{0.6}{\ampere\per\watt}  \\ 
 2019 & \cite{Benedikovic2019} & p-n (Ge) & \SI{1550}{\nano\metre} & \SI{30}{\giga\hertz} & \SI{150}{\nano\ampere} & \SI{0.84}{\ampere\per\watt}\\
 2019 & \cite{Chatterjee2019} & p-i-n (Si) & \SI{850}{\nano\metre} & \SI{14}{\giga\hertz} & \SI{75}{\nano\ampere} & \SI{0.44}{\ampere\per\watt} \\
 2020 & \cite{Chatterjee2020} & MSM (Si) & \SI{850}{\nano\metre} & \SI{7.5}{\giga\hertz} & \SI{292}{\pico\ampere\per\micro\metre\squared} & \SI{0.81}{\ampere\per\watt} \\
 2021 & \cite{lischke2021ultra} & p-i-n (Ge) & \SI{1550}{\nano\metre} & \SI{265}{\giga\hertz} & \SI{200}{\nano\ampere} & \SI{0.3}{\ampere\per\watt} \\
 2021 & \cite{Yanikgonul2021} & p-n (Si) & \SI{685}{\nano\metre} & \SI{19.1}{\giga\hertz} & \SI{120}{\nano\ampere} & \SI{0.83}{\ampere\per\watt} \\
 2022 & \cite{Lin2022} & p-n (Si) & \SI{488}{\nano\metre} & \SI{9}{\giga\hertz} & \SI{226}{\pico\ampere} & -- \\
 2022 & \cite{Cuyvers2022} & p-i-n (Si) & \SI{850}{\nano\metre} & \SI{6}{\giga\hertz} & \SI{107}{\pico\ampere} & \SI{0.19}{\ampere\per\watt} \\
 2022 & \cite{Chatterjee2022} & MSM (Si) & \SI{850}{\nano\metre} & -- & \SI{2.9}{\nano\ampere\per\micro\metre\squared} & \SI{0.56}{\ampere\per\watt} \\
 2024 & \cite{Xue2024} & p-i-n (Ge) & \SI{1336}{\nano\metre} & \SI{40}{\giga\hertz} & 1-1000\si{\pico\ampere} & \SI{0.29}{\ampere\per\watt}\\
 2024 & \cite{Peng2024} & p-i-n (Si) & \SI{1309}{\nano\metre} & \SI{40}{\giga\hertz} & \SI{1}{\nano\ampere} & \SI{0.4}{\ampere\per\watt} \\
 \hline \hline
 \multicolumn{7}{|c|}{\textbf{Integrated homodyne detectors}} \\
 \hline\hline
 Year & Ref. & Type & $\lambda$ & Bandwidth & Max. clearance & CMRR  \\
 \hline\hline
 2018 & \cite{raffaelli2018homodyne} & p-i-n (Ge) & \SI{1550}{\nano\metre} & \SI{150}{\mega\hertz} & \SI{11}{\decibel} & \SI{28}{\decibel}  \\
 2019 & \cite{zhang2019integrated} & p-i-n (Ge) & \SI{1550}{\nano\metre} & \SI{10}{\mega\hertz} & \SI{5}{\decibel} & \SI{25}{\decibel}  \\
 2021 & \cite{tasker2021silicon} & p-i-n (Ge) & \SI{1550}{\nano\metre} & \SI{1.7}{\giga\hertz} & \SI{14}{\decibel} & \SI{52}{\decibel}  \\
 2021 & \cite{bruynsteen2021integrated} & p-i-n (Ge) & \SI{1550}{\nano\metre} & \SI{1.5}{\giga\hertz} & \SI{28}{\decibel} & \SI{80}{\decibel}  \\
  2022 & \cite{milovanvcev2022chip} & p-i-n (InGaAs) & \SI{1550}{\nano\metre} & \SI{2.6}{\giga\hertz} & \SI{21}{\decibel} & \SI{50}{\decibel}  \\
 2023 & \cite{Jia2023} & p-i-n (Ge) & \SI{1550}{\nano\metre} & \SI{1}{\mega\hertz} pulsed & \SI{19}{\decibel} & \SI{87}{\decibel}  \\
 2024 & \cite{Ng2024} & p-i-n (Ge) & \SI{1550}{\nano\metre} & \SI{4.75}{\giga\hertz} & \SI{12.9}{\decibel} & \SI{39.6}{\decibel}  \\
 2024 & \cite{tasker2024bi} & p-i-n (Ge) & \SI{1550}{\nano\metre} & \SI{15.3}{\giga\hertz} & \SI{12}{\decibel} & \SI{27}{\decibel}  \\ 
 2025 & \cite{bian202520} & p-i-n (Ge) & \SI{1550}{\nano\metre} & \SI{2}{\giga\hertz} & \SI{10}{\decibel} & -  \\ \hline
 \end{tabular}
 \caption{High-bandwidth integrated detectors. Top: integrated photodetectors; bottom: integrated homodyne detectors. }
\label{table:detectors}
\end{table*}
Another limitation was that the photodiode balancing depended solely on the static MMI splitting ratio, as opposed to utilising an integrated tunable MZI, ultimating limiting the detector's CMRR to 28dB.

Three years later in 2021, Tasker \textit{et al.} designed an SOI PIC featuring an integrated homodyne detector with Ge p-i-n photodiodes fabricated by IMEC \cite{tasker2021silicon}. The most significant change was the direct wire-bonding of the PIC to an unpackaged off-the-shelf TIA die, circumventing the parasitic capacitances introduced by discrete electronic packaging and PCB traces. This change extended the 3dB bandwidth of the detector to 1.7 GHz, with shot noise clearance up to 9 GHz. This chip also featured integrated phase shifters such that the CMRR could be carefully optimised using an integrated MZI, with a reported CMRR of $>$52dB.

The benefits of reducing parasitic capacitances by wire-bonding directly to electronics was also demonstrated in 2021 by Bruysteen \textit{et al.} at IMEC technologies \cite{bruynsteen2021integrated}. A similar 3dB bandwidth of 1.5GHz was achieved, and by designing their own custom ultra-low noise TIA, a shot noise clearance of 28dB was demonstrated. While \cite{bruynsteen2021integrated} and \cite{tasker2021silicon} consider the inverse relationship between bandwidth and capacitance, a 2024 design from Si Qi Ng \textit{et al.}  \cite{Ng2024} also accounts for the inverse relationship between bandwidth and resistance. By wirebonding the output of the balanced photodiodes directly onto a thin-film \SI{50}{\ohm} resistor and RF amplifier on a PCB, the \SI{3}{\decibel} bandwidth was enhanced to \SI{4.75}{\giga\hertz} at the cost of a slightly reduced shot noise clearance. In the same year, an integrated homodyne detector based on InGaAs photodiodes and a TIA die was demonstrated by Bai \textit{et al.} with hybrid packaging \cite{bai202118}. While InGaAs can exhibit higher speed photodetection than silicon \cite{Serikawa2018}, heterogeneous integration of III-Vs is currently a much less mature and more expensive technology \cite{feng2012epitaxial}. A recent work from Bian \textit{et al.} \cite{bian202520} demonstrates a balanced detector with a shot noise clearance of \SI{10}{\decibel} and a bandwidth of over \SI{2}{\giga\hertz}. By packaging the fibre array directly onto the grating coupler, an overall chip loss of \SI{4.45}{\decibel} is achieved, and a real-time QRNG generation of 20Gbps is demonstrated.

While the bandwidths of \cite{tasker2021silicon}, \cite{bruynsteen2021integrated} and \cite{Ng2024} have significantly improved upon the earlier designs, they are still ultimately limited by the parasitic capacitance introduced by the wirebond connecting the photonic chip to the electronics. By utilising foundry-fabricated monolithic electronic PIC (ePIC) devices and removing the bonding and packaging capacitances, Tasker \textit{et al.} \cite{tasker2024bi} demonstrated a \SI{3}{\decibel} bandwidth of \SI{15.3}{\giga\hertz}.

\section{Experimental challenges for monolithic integration}
\label{sec:OnChipIntegrated}
Integrating CV experiments places stringent requirements on the optical fields for both squeezed light generation and detection. When implementing experiments requiring squeezing careful thought must be given to preparing the optical fields required to both generate and detect the state, with these requirements being dependent on the type of squeezing desired. Whilst the direct or balanced detection of bright and intensity difference squeezing requires only photodiodes, the detection of quadrature squeezing requires homodyne detection, and thus an LO. 
For SMSS the LO requirement is a single spectral tone. It is also possible to detect TMSS with a single LO using sideband detection by examining the photocurrent at the beat frequency equal to the detuning of TMSS from the LO. With on chip Homodnye detection limited to a bandwidth of tens of \si{\giga\hertz}, this detuning must be sub-nanometre to create a measureable beat-note. With SFWM based sources suffering from noise processes close to the pump it is common to generate and detect squeezing at disparate wavelengths, ruling out sideband detection and instead requiring a bichromatic LO \cite{Marino2007Birchomatic}. 

As homodyne detection is an interferometric measurement, phase coherence between the LO and squeezing is essential. This usually satisfied by ensuring  phase coherence between the LO and pumps for the generation, since the phase of the squeezing is dictated by the phase of the pump. For a SMSS generated by SPDC the pump and LO can be generated from a single source at the LO frequency and performing SHG to generate the pump. With FWM, the challenge is amplified by the requirement for a bichromatic pump and single LO, or single pump and bichromatic LO. Vaidya \textit{et al.} achieved this by phaselocking three independent lasers to pump a Si\textsubscript{3}N\textsubscript{4} microring resonator to generate squeezing, and produce the required bichromatic LO for its detection \cite{vaidya2020broadband}. A similar approach reduced the complexity by phaselocking two independent sources together and generating the other by stimulated four wave mixing in nonlinear fibre \cite{zhang2021squeezed}. The squeezed frequency comb demonstrated by Yang \textit{et al.} \cite{yang2021squeezed} required a bichomatic LO, generated by electro-optically modulating the pump using an external LiNbO\textsubscript{3} phase modulator. A programmable optical filter was employed to select sidebands for the LO, and hence which TMSS was interrogated. 
 
Work from the Lipson group \cite{zhao2020near} demonstrated a device where both a SMSS and an LO were generated on-chip using two identical ring resonators with different coupling conditions. A dual pump scheme was used with one ring operating below threshold to generate single mode quadrature squeezing, with the other operating above threshold to generate the LO. In this work, \SI{0.8}{\decibel} of quadrature squeezing was measured, with \SI{3.09}{\decibel} inferred after correcting for losses and detection efficiency, and an estimated maximum of \SI{3.5}{\decibel}. 

This represents a growing trend to integrate all aspects of the generation, manipulation and detection of squeezing onto the same chip. There is, however, a disparity between the platforms best suited to high-quality sources of squeezing and those for which high efficiency quantum-limited homodyne detectors have been demonstrated. Although Si\textsubscript{3}N\textsubscript{4} has shown great promise in the generation of squeezing it is challenging to integrate telecom-compatible photodetectors meaning that, to date, integrated homodyne detectors have solely been demonstrated on the Si platform. Whilst high bandwidth, waveguide-coupled Ge photodiodes have been demonstrated on Si\textsubscript{3}N\textsubscript{4} their implementation has so far only been demonstrated in the higher loss PECVD platform \cite{lischke2019Silicon}. 
Access to photodetectors on Si is also possible by moving away from the telecom infrastructure to visible wavelengths. This comes with the drawback of higher linear losses due to scattering at the waveguide interfaces and nonlinear loss by TPA as the two-photon energy approaches that of the bandgap. To circumvent these platform limitations, hybrid and heterogeneous integration is being actively explored where the microtransfer printing of InP-InGaAs photodiodes onto Si\textsubscript{3}N\textsubscript{4} waveguides has been demonstrated \cite{Qin2024Microtransfer}. Flip-chip bonding photodiode modules to a separate photonic chip has been explored in CV applications \cite{bai202118} to allow the interfacing of InGaAs photodiodes with a QRNG on an SOI platform. Whilst permitting a compact design, the grating coupler losses are comparable to coupling off-chip. Although possessing mature photodiodes, the use of SOI in integrated CV experiments is waning due to the deleterious linear and nonlinear losses in Si that suppress the generated squeezing. To obtain the squeezing levels demanded by many applications, the overall circuit loss must be reduced. Monolithic integration of sources and detectors on a single platform removes the major loss contributor of coupling squeezing off chip, motivating research into the integration of homodyne detectors onto other platforms.

\section{Applications, outlook and conclusions}
\label{sec:Conclusions}

\subsection{Applications to quantum technologies}
\label{ssec:Applications}

\begin{figure*}
    \centering
    \includegraphics[width=0.8\linewidth]{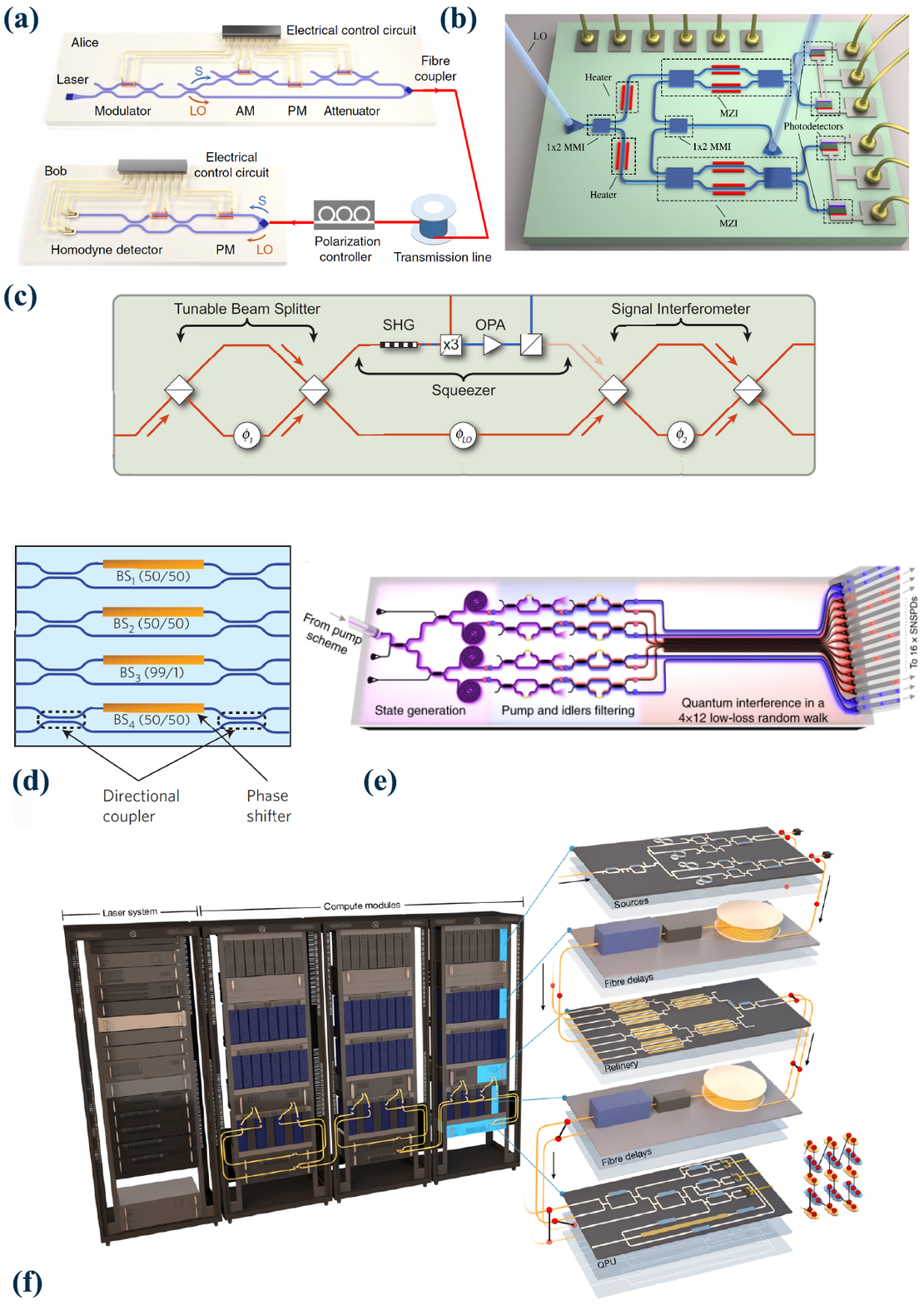}
    \caption{Integrated devices demonstrating experiments towards quantum technologies using continuous-variable regimes. (a) A CV-QKD transmitter and receiver reported by Zhang \textit{et al.} \cite{zhang2019integrated}. (b) An integrated photonic-electronic receiver for CV-QKD reported by Hajomer \textit{et al.} \cite{hajomer2024continuous}. (c) An optical phase sensor reported by Stokowski \textit{et al.} \cite{stokowski2023integrated}. (d) The first demonstration of deterministic CV entanglement generation on-chip, reported by Masada \textit{et al.} \cite{masada2015continuous}. (e) A PIC for executing quantum algorithms, first demonstrated by Paesani \textit{et al.} \cite{Paesani2019} (f) A schematic of the rack-mounted system Aurora from Xanadu \cite{aghaee2025scaling}.}
    \label{fig:Applications}
\end{figure*}

The advantages of a CV QKD system includes the ability to produce high secret key rates, as the information is encoded in continuous variables \cite{jouguet2014high}. Recently, a number of high-rate CV-QKD systems have been engineered which include both Gaussian \cite{Gaussian2,Gaussian3} and discrete-modulation cases \cite{Discrete1}. These systems are well-suited for photonic integration due to compatibility with standard telecom components. The first proof of principle chip-based CV QKD system was demonstrated in Zhang \textit{et al.} \cite{zhang2019integrated}. Figure \ref{fig:Applications}(a) shows the experimental setup of the system, which uses integrated silicon PICs to demonstrate a Gaussian-modulated coherent state CV-QKD protocol. A \SI{1550}{\nano\metre} laser is coupled onto the transmitter chip and split into two paths with a 1:99 ratio, for the signal and LO respectively. The signal path was modulated by a set of amplitude and phase modulators on the sideband ranging from 1–10 MHz. After the signal is transmitted over the quantum channel, the receiver arbitrarily measures quadratures with homodyne detection and filters out the required frequency, with secret key rate of \SI{0.25}{\mega\bit\per\second} reported.
Recent work demonstrated a significant improvement in the secret key rate \cite{hajomer2024continuous} by improving the receiver's bandwidth. Specifically, a co-integrated phase diverse receiver was introduced, consisting of a silicon photonics optical front-end and custom-integrated trans impedance amplifiers (TIAs) designed in a 100 nm GaAs pHEMT technology. Figure \ref{fig:Applications} (b) depicts a schematic of the receiver system. In contrast to Zhang \textit{et al.}, this work uses a discrete constellation space for modulation, making it more compatible with high-speed wireline components. Operating at a classical telecom symbol rate of 10 GBaud, they have reported that their QKD system generates high secret key rates - exceeding \SI{0.7}{\giga\bit\per\second} over a \SI{5}{\kilo\metre} distance and \SI{0.3}{\giga\bit\per\second} over a \SI{10}{\kilo\metre} distance. At a similar time, Bian \textit{et al.} reported a CV QKD system with secret key rates of \SI{1.3}{\mega\bit\per\second} over a \SI{28.6}{\kilo\metre}. The receiver chip was a packaged homodyne detector with a \SI{1.5}{\giga\hertz} bandwidth and tuneable balancing, permitting good suppression of excess noise and allowing higher transmission distances.

Quantum metrology provides a route to surpass classical limits imposed on precision by exploiting the properties of quantum states \cite{giovannetti2006quantum}. The intrinsic sub shot noise characteristics of squeezed light has been a valuable instrument in enhancing the resolution \cite{taylor2014subdiffraction} or sensitivity \cite{qin2023unconditional} of measurements beyond the standard quantum limit. Quantum-enhanced sensing can be demonstrated by simply probing a sample with squeezed light and using a homodyne detector to measure the changes induced by the properties of the sample \cite{atkinson2021quantum}. Therefore integrated photonics provides a route to developing compact, repeatable and robust quantum photonic sensors. Achieving full monolithic integration of the detectors and squeezing source allows the quantum noise properties to be fully utilised, avoiding losses from coupling on or off chip.

A group at Stanford led by Safavi-Naeini recently demonstrated an integrated optical phase sensor \cite{stokowski2023integrated} employing a squeezed vacuum state on-chip. The device was based on two consecutive PPLN waveguides for SHG followed by parametric amplification, integrated alongside tunable phase shifters and beamsplitters. Figure \ref{fig:Applications}(c) shows a schematic of the PIC layout. The OPA waveguide acts as an optical phase sensor, with $2.7$\% noise reduction measured over a long timescale measurement of \SI{13}{\second} and a total of \SI{1.5}{\decibel} inferred on-chip. The squeezed state is used for the quantum-enhanced sensing of an RF signal, demonstrating a total SNR improvement of 3.7\%. Where the majority of loss experienced by the squeezed vacuum arises from coupling the light off-chip for detection, integration of the homodyne detector would allow for further SNR improvement.

CV-QKD and quantum sensing can be demonstrated with a Gaussian source and homodyne detection. To build a universal CV quantum computer, an element of non-Gaussianity is required: for example, performing detection with single photon detectors or utilising non-Gaussian CV states, which typically require single photon detectors for generation\cite{Walschaers2021}. Therefore the full monolithic integration of a CV system for quantum computing presents the additional challenges of cryogenic compatibility and heterogeneous integration.

Despite these complexities, integrated photonics has already been explored in CV quantum computing systems. One route to a practical quantum computational advantage is generating large-scale cluster states by exploiting the deterministic entanglement benefitted from squeezed light. The first step of building these %large scale CV 
cluster states is to construct CV-EPR states by interfering states squeezed along orthogonal quadratures. This deterministic entanglement process was demonstrated in 2015 utilising a set of integrated MZIs to interfere the squeezed states, and then to interfere the CV-EPR states with the LO for homodyne detection \cite{masada2015continuous}. Figure \ref{fig:Applications}(d) shows a schematic of the PIC used for this. Here, both the detection and squeezing generation were performed off-chip, such that while the inseparability criteria was satisfied, the entanglement generated was ultimately limited by coupling losses.

In 2019 a valuable use of integrated photonics and squeezed light on the path to a scalable and useful photonic quantum computer was demonstrated by Paesani \textit{et al.} \cite{Paesani2019}. They demonstrate Gaussian boson sampling (GBS) on a silicon photonic chip, depicted in Figure \ref{fig:Applications}(e) generating spatial distribution samples of four photons over twelve spatial modes. While the photons are detected off-chip using SNSPDs, they are derived from four weak SMSS generated at four independent SFWM spiral sources, using a dual-pump scheme. The pump and heralding photon filtering, and the linear optics interference network required for GBS are implemented are the same chip as the sources. The GBS circuit is then mapped to the molecular vibronic spectra of a synthetic molecule, demonstrating the potential of integrated photonics for quantum simulations.

Work by Arrazola \textit{et al.} \cite{arrazola2021quantum} implemented four integrated sources of squeezing \cite{vaidya2020broadband} to demonstrate three quantum algorithms on chip. Four independent TMSS states were able to be arbitrarily routed on a user-programmable gate sequence over four spatially separate waveguide modes. Using photon-number-resolving detectors, the full eight-mode Gaussian state was measured in the Fock basis, with an estimated effective squeezing of \SI{8}{\decibel}. The chip was used to demonstrate GBS, simulate vibronic spectra of chemicals and perform a graph similarity experiment. This work was extended by the same company in a very recent demonstration of a quantum computer `scale-model' of $35$ photonic chips able to generate an entangled cluster state \cite{aghaee2025scaling}. Figure \ref{fig:Applications}(f) contains a schematic of the rack-mounted Aurora system.
The design is broken down into three stages each utilising a separate PIC composition. First a Si\textsubscript{3}N\textsubscript{4} chip is used to generate squeezing and probabilistically generate non-Gaussian states from GBS. These states, or the squeezed state included in the case when the GBS protocol is unsuccessful in a given clock cycle, are fed into a thin-film Lithium Niobate chip, chosen for its fast electro-optic switching capabilities, to route and entangle these states. As the states generated by GBS are precursors to the target states for QIP, in this step many copies of these states would then be refined into a state with high fidelity to the target. However, this refining was not performed in this work instead using these precursor states in the final cluster. A final `QPU' chip is used to create clusters and measure cluster states, relying on a Si\textsubscript{3}N\textsubscript{4} waveguide layer for manipulation and a Si layer with waveguide coupled Ge PDs to form HDs. It is worth noting that the loss incurred in the design is well above that budgeted for fault tolerance, mostly limited by the coupling losses between chips. While this design is a way off performing useful quantum enhanced computation, it demonstrates the potential of a modular structure that utilises different integrated photonic platforms for their corresponding advantages. Where only a portion of the structure needs to be kept at cryogenic temperatures to implement the required non-Gaussianity via photon number resolving detectors, this design demonstrates the relative energy efficiency of the continuous variable paradigm for quantum information. 
%Quantum computing papers
%https://www.nature.com/articles/s41586-022-04725-x
%https://www.nature.com/articles/s41567-019-0567-8#Sec3
%https://www.nature.com/articles/s41586-024-08406-9#MOESM1

\subsection{Outlook and conclusions}
\label{ssec:Final}
The use of CV as a platform for realising photonic quantum technologies has elicited widespread interest in the research community for a variety of end applications. The ability to deterministically generate quantum states at room temperature with linear optics is an appealing aspect when considering the desire to scale-up and deploy quantum technologies. Implementing techniques based on the mature classical communications field, the detection of CV states on chip has been demonstrated recently and is progressing rapidly.  Advances in the last decade have outlined a blueprint for fabricating these devices, both sources and detectors, with CMOS-compatible processes. These developments offer promise for solving the remaining engineering challenges of monolithic integration. The remaining questions for the integrated CV community appear to rest ultimately on the choice of material to achieve, simultaneously on a monolithic chip, the highest levels of squeezing, and the most efficient and fast detectors for its measurement. The aspirations of fully-integrated CV experiments on CMOS-compatible ePICs can expedite a wide array of real-world quantum technologies, including quantum sensing, communication and computing.

\small
\subsection{Acknowledgements}

This was supported by European Research Council starting grant ``PEQEM'' (ERC-2018-STG 803665) and EPSRC grant ``Mono-Squeeze'' (EP/X016218/1, EP/X016749/1).
B.P. \& O.M.G. acknowledge support from EPSRC Quantum Engineering Centre for Doctoral Training (EP/S023607/1).
J.C.F.M is grateful for support from his Philip Leverhulme Prize.

\bibliography{apssamp}% Produces the bibliography via BibTeX.

\end{document}